\documentclass[pra,twocolumn,showpacs]{revtex4-2}
\usepackage{orcidlink}
\pdfoutput=1

\usepackage{graphicx}
\usepackage{amssymb}
\usepackage{mathtools}
\usepackage{epstopdf}
\usepackage{physics}
\usepackage{dsfont}
\usepackage{amsmath}
\usepackage{hyperref} 
\usepackage{xcolor}
\usepackage[normalem]{ulem}
\usepackage{titlesec}

\newcommand{\Bra}[1]{\langle \! \langle#1|}
\newcommand{\Ket}[1]{|#1\rangle \! \rangle}

\newcommand{\KetBra}[2]{{\Ket{#1}\Bra{#2}}}
\newcommand{\uv}{\text{\tiny UV}}
\newcommand{\ir}{\text{\tiny IR}}
\newcommand{\mix}{\text{\tiny MIX}}

\graphicspath{ {./figures/} }

\begin{document}

\title{Effective dynamics from minimising
  dissipation}

\author{Antonio F.\ \surname{Rotundo} \orcidlink{0000-0002-5251-5693}}
\email[]{antoniofrancesco.rotundo@unipv.it}
\affiliation{Dipartimento di Fisica
  dell'Universit\`a di Pavia, via Bassi 6, 27100
  Pavia} \affiliation{Istituto Nazionale di Fisica
  Nucleare, Gruppo IV, via Bassi 6, 27100 Pavia}
\author{Paolo \surname{Perinotti}\orcidlink{0000-0003-4825-4264}}
\email[]{paolo.perinotti@unipv.it}
\affiliation{Dipartimento di Fisica
  dell'Universit\`a di Pavia, via Bassi 6, 27100
  Pavia} \affiliation{Istituto Nazionale di Fisica
  Nucleare, Gruppo IV, via Bassi 6, 27100 Pavia}
\author{Alessandro \surname{Bisio}\orcidlink{0000-0002-9356-3448}}
\email[]{alessandro.bisio@unipv.it}
\affiliation{Dipartimento di Fisica
  dell'Universit\`a di Pavia, via Bassi 6, 27100
  Pavia} \affiliation{Istituto Nazionale di Fisica
  Nucleare, Gruppo IV, via Bassi 6, 27100 Pavia}

\date{\today}

\begin{abstract}
  It is known that the same physical system can be
  described by different effective theories
  depending on the scale at which it is observed.
  In this work, we formulate a prescription for
  finding the unitary that best approximates the
  large scale dynamics of a quantum system
  evolving discretely in time, as it is the case
  for digital quantum simulators.  We consider the
  situation in which the degrees of freedom of the
  system can be divided between an IR part that we
  can observe, and a UV part that we cannot
  observe.  Following a principle of minimal
  dissipation, our goal is to find the unitary
  dynamics that best approximates the (generally
  non unitary) time evolution of the IR degrees of
  freedom.  We first prove that when the IR and UV
  degrees of freedom are weakly coupled, the
  unitary that maximises the fidelity is given by
  a mean-field dynamics and the error is given by a sum of energy variances.
We then apply our results
to a one dimensional quantum walk, which
  is known to reproduce the Dirac equation in the
  small mass and momenta limit.  We find that in
  this limit the effective IR dynamics is obtained
  by a mass redefinition.
\end{abstract}


\maketitle

\emph{Introduction.}  Effective theories are a
powerful and efficient way of organising our
description of physical processes
\cite{polchinski1992effective, 10.1007/BFb0104294,
  kaplan2005five, burgess2007introduction} The rough idea is that, in order to
describe physical phenomena at some energy scale $E$, one does
not need a full understanding of physics at much
larger energy scales $\lambda \gg E$.  The high-energy
degrees of freedom approximately decouple, allowing one to describe the low-energy dynamics through an
effective unitary theory.  The only role of the high-energy degrees of freedom 
%theory
%is only required 
is to determine the parameters of this
effective theory.
The challenge of determining the effective theory
has been studied for Hamiltonian dynamics,
particularly within field theories \cite{wilson1971renormalization, polchinski1984renormalization, georgi1991shell, polchinski1992effective, 10.1007/BFb0104294,kaplan2005five, burgess2007introduction, henning2016use}.
However, the case of system evolving in discrete
time remains largely unexplored. The goal of this
article is to help bridge this critical gap. 
%We identify two main reasons t.  

This kind of analysis is thus relevant 
for physical models such as
%of interest that inherently evolve in
%discrete steps, most notably 
quantum walks (QW)
\cite{portugal2013quantum, aharonov2001quantum,
  ambainis2003quantum, venegas2012quantum} and
quantum cellular automata (QCA)
\cite{farrelly2020review, arrighi2019overview,
  schumacher2004reversible}.
QCAs and QWs are universal models of computation equivalent to the quantum circuit model \cite{watrous1995one, childs2009universal}.
In particular, QWs are
interesting both as a quantum computing primitive,
especially in search algorithms \cite{portugal2013quantum,
  aharonov2001quantum, ambainis2003quantum,
  venegas2012quantum}, and for quantum
simulation \cite{arrighi2014dirac, arnault2016quantum, arnault2019discrete, di2012discrete}.
QCAs describe quantum systems evolving under unitary and strictly local dynamics \cite{arrighi2011unitarity}, so they can be naturally used for quantum simulations \cite{arrighi2020quantum, farrelly2020discretizing}.
They are also the appropriate framework to handle Floquet systems \cite{po2016chiral}.
It is natural to ask what kind of effective dynamics these models generate at large distances. 
More generally discrete-time dynamics is a typical framework
for quantum simulation, in which the target dynamics is obtained by 
Trotterisation \cite{lloyd1996universal, zalka1998efficient, jordan2012quantum, georgescu2014quantum, childs2021theory, bauer2023quantum}.
%, which stands as one of the most
%promising applications of quantum computing.  
%Most
%quantum hardware platforms operate in discrete
%time steps, which means that, in practice, also for Hamiltonian dynamics only a
%discrete-time approximation of the target dynamics
%can be implemented.  
%Typically time is  discretised by Trotterisation.
Therefore, one could be interested in what is the effective dynamics of the simulator when small-scale details of the implementation, which cannot be perfectly tuned, are traced out.

%Often we are interested only in reproducing the target dynamics at a coarse level
%The error introduced by Trotterisation step are well characterised \cite{, childs2019nearly}, and so one can make sure that the simulation is close to the target dynamics.
%However, often we are interested only in simulating 
%This certainly implies that the simulation implements the target effective theory at large scales.
%However, as we said, the effective theory should be insensitive to the precise microscopic implementation, and so requiring that the simulation is close to the microscopic theory might very well be an overshot. 
%What we need is only that the effective dynamics generated by the simulator at large scale is close to the target effective dynamics.
%Therefore, developing the tools to derive and analyse effective theories directly for discrete-time dynamics would be very valuable for quantum simulation.

When we are interested in simulating quantum phenomena which are well described by some effective theory, we face two options.
The first one is to directly simulate the effective theory on our quantum hardware, see for example \cite{dumitrescu2018cloud, roggero2020quantum, bauer2021simulating, watson2023quantum}; the second one is to simulate a more fundamental theory known to reproduce the effective physics of interest at a large scale, see for example \cite{osborne2021quantum, osborne2023conformal}.
One advantage of this second approach is that by construction the quality of the simulation is largely insensitive to the details of the microscopic implementation, which is especially valuable for current pre-fault tolerant quantum computers \cite{preskill2018quantum}. Typically, one can figure out 
%To implement this 
a microscopic theory by time-discretising the target theory, recurring to Trotterisation.
This approach allows for a high control, since its approximation errors are well characterised \cite{childs2021theory, childs2019nearly}.
%, and so one can make sure that the simulation is close to the target microscopic theory.
%This certainly implies that the simulation implements the target effective theory at large scales.
%However, as we said, the effective theory should be insensitive to the precise microscopic implementation, and so requiring that the simulation is close to the microscopic theory might very well be an overshot. 
%What we need is only that the effective dynamics generated by the simulator at large scale is close to the target effective dynamics.
On the other hand, if we rely on universality---i.e.~the irrelevance of microscopic details for the large-scale behaviour---it should be in principle possible to resort to a much simpler, yet highly effective microscopic model. Therefore, there is room for simplification of implementations of a Trotterised model. Developing the tools to derive and analyse effective theories directly for discrete-time dynamics would be very valuable to validate the simplifications that one might consider.

Let us then consider a quantum system with Hilbert space
$\mathcal{H}$, evolving in discrete steps by the
repeated action of a unitary $U$.  A qualitative
difference with Hamiltonian dynamics is that,
knowing only $U$, we don't have a well-defined
notion of energy.  We assume that the Hilbert
space can be split between a fine-grained (UV)
part and a coarse-grained (IR) part,
$\mathcal{H}=\mathcal{H}_{\uv}\otimes
\mathcal{H}_{\ir}$, of which we can only observe
the latter.  Then all the measurable phenomenology 
%our possible measurements are 
is determined by the IR reduced density matrix
$\rho_{\ir}=\Tr_{\uv}\rho$, where $\rho$ is the
state of the whole system.  In general, the
dynamics in the IR sector of the theory is
non-unitary, because $U$ produces entanglement
between IR and UV degrees of freedom.  We then address
the following question: what is the unitary
$U_{\ir}$ that best approximates the dynamics of
the IR sector?  In other words, our goal is
finding the unitary $U_{\ir}$ such that
\begin{equation}
\begin{gathered}
\includegraphics[width=0.35\textwidth]{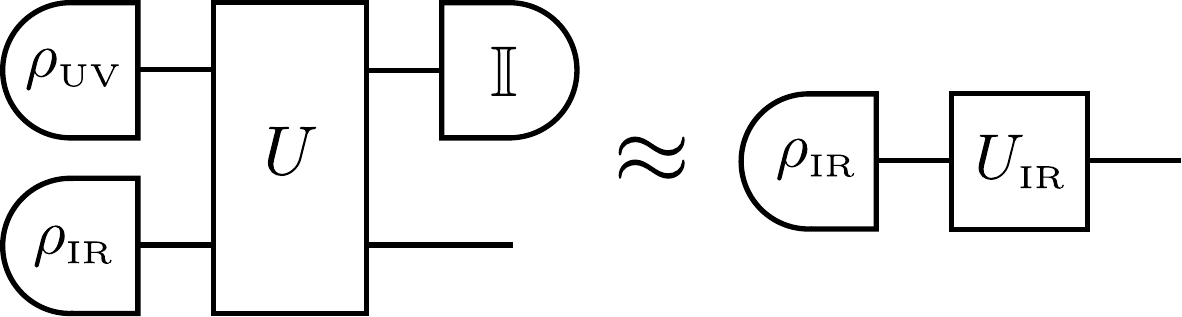}
\end{gathered}\quad.
\end{equation}
Once we know that $U_\ir$ is a good approximation
of the dynamics at the scale of interest, we can
use it in place of $U$.  The benefit of allowing
for a unitary approximation rather than using the
exact non-unitary channel is that it simplifies
both the analytical study and the simulation of
the dynamics.  At a more fundamental level,
understanding which models of discrete dynamics
lead to effectively unitary theories at long
distances is important to understand whether they
could be viable UV completion of known quantum
field theories \cite{PhysRevA.90.062106,
  farrelly2014discrete}.
 
As a criterion to quantify the quality of the
approximation we use the channel fidelity
\cite{raginsky2001fidelity}, a well-established
measure of distinguishability between quantum
channels.  Moreover, this measure of distance is
also justified by drawing a parallel with an
analogous problem in the physics of open quantum
systems.  In this context, the UV degrees of freedom that we trace
out can be viewed as an external environment,
which we cannot control or observe.  Open quantum
systems evolve according to a master
equation, which can be diveided into a dissipative
part and a Hamiltonian (or unitary) part.  Although this division 
is not unique, it is tipically chosen so that
the dissipative term takes Lindblad form.  In a
recent work \cite{hayden2022canonical}, it was
shown that this seemingly ad-hoc division
corresponds
to selecting the smallest possible dissipative term
according to a metric that reduces to the channel
fidelity when one of the channels
is unitary.  This approach has been 
epitomsed in Ref. \cite{colla2022open} as the
\emph{principle of minimal dissipation}. From
this perspective, we can say that the effective
IR dynamics that we consider is the one minimising
dissipation.

Finding the best unitary $U_\ir$ by maximising the
channel fidelity is, in general, a challenging task,
and we do not expect a good approximation
to be achievable for arbitrary unitaries $U$.
Therefore, we focus on the case in which the IR
and UV degrees of freedom are weakly coupled. 
To leading order in the coupling, we show that the solution
is a mean-field dynamics.  We then apply
this result to a QW on $\mathbb{Z}$, where the spatial
lattice on which the walk takes place provides a
natural way of splitting the Hilbert space between
a UV and a IR part.  Namely, we group the lattice
points in bins: the IR degree of freedom specifies
the bin in which the walker is located, the UV
degree of freedom specifies the position within
the bin.  In particular, we consider the Dirac QW
of \cite{Bisio2015244}, which is known to
reproduce the Dirac equation in the small mass and
large wavelength limit.  We find that in this same
limit the effective dynamics is given by a simple
rescaling of the mass, that encapsulates all the
dependence on the UV degrees of freedom.

Before continuing, we briefly point out previous
works that have already considered the
renormalisation of systems evolving in discrete
time steps.  The question of how to renormalise QW
was considered in
\cite{boettcher2014renormalization,
  boettcher2017renormalization}. The main
difference with our work is that there the authors
renormalise the QW at the level of the generating
function $\tilde{\psi}(x,z)=\sum_t \psi(x,t)z^t$,
in which the explicit dependance on time is
eliminated. Here, $\psi(x,t)$ is the wavefunction
of the walker.  Some form of
coarse-graining for QW was also considered in
\cite{arrighi2014discrete} as part of an attempt
at defining discrete Lorentz invariance.  
Coarse-graining of quantum cellular automata (QCA) has
been considered in \cite{duranthon2021coarse} and
\cite{trezzini2024renormalisation}.  In the first,
the authors consider a special type of QCA and
implement a coarse graining procedure tailored for
it. The prescription considered in this work
should instead be generally applicable.  In the
second, the authors considered a general
coarse-graining scheme for QCA, but imposed the
requirement that the coarse grained dynamics is
\emph{exactly} unitary. In this work, we consider
the more general situation in which the
coarse-grained dynamics is only approximately
unitary.

The rest of the article is organised as follows.
We begin by explaining our prescription for
finding the effective IR dynamics.  We then
specialise to the case in which the UV and the IR
are only weakly coupled and find the effective
dynamics to leading order in this coupling.  We
apply this result to the Dirac QW on the line, and
find that when the coupling between UV and IR is
weak, the coarse-graining reduces to a mass
renormalisation.  We conclude by pointing out some
possible extensions of this work.

\emph{Effective unitary dynamics.}
Let us
denote with
$\rho=\rho_{\ir}\otimes\rho_{\uv}$  the initial state
of the system, where $\rho_{\ir} \in \mathcal{L}(\mathcal{H}_{\ir})$ and $\rho_{\uv} \in \mathcal{L}(\mathcal{H}_{\uv})$.
%  already presents strong entanglement
% between IR and UV degrees of freedom, our unitary
% approximation is bound to fail. So, we assume
% that the initial state of the system is a product
% state.
% We will consider the case in which 
% the Hilbert space $\mathcal{H}=\mathcal{H}_{\ir}\otimes
% \mathcal{H}_{\uv}$ is finite dimensional, i.e. $d_{\uv} \coloneqq \dim \mathcal{H}_{\uv} < \infty$ and
% $d_{\ir} \coloneqq \dim \mathcal{H}_{\ir} < \infty$.
The channel describing the evolution of
$\rho_{\ir}$ depends both on the specific system
we consider and on $\rho_\uv$.
We will
keep $\rho_\uv$ generic, and only specify it when we consider
examples.
If we had some notion of energy for the
UV system,
a reasonable choices would be to take 
$\rho_\uv$  to be the
thermal state at some given temperature.

For a fixed $\rho_\uv$, the one-step evolution of the IR state $\rho_{\ir}$
is described by the channel $\mathcal{D}$ defined as follows:
\begin{align}
  \label{eq:2}
\mathcal{D}(\rho_{\ir}) \coloneqq \Tr_{\uv}\bigl[U(\rho_{\ir}\otimes \rho_{\uv}) U^\dagger\bigr].
\end{align}
Our goal is to find the unitary channel
$\mathcal{U}_{\ir}(\rho) \coloneqq U_{\ir} \rho
U^\dag_{\ir}$ which is as close as possible to
the channel $\mathcal{D}$.

Let us consider the case in which the Hilbert
space of the IR degrees of freedom is finite
dimensional 
$d_{\ir} \coloneqq \dim \mathcal{H}_{\ir} <
\infty$.  
The latter simplifying assumption can be justified by the fact that, in practice, only a finite number of degrees of freedom will be experimentally accessible. On formal grounds, this  could be achieved by projecting down to a finite dimensional subspace of the original Hilbert space or considering periodic boundary conditions.
As a measure of distance we
use the channel fidelity
\cite{raginsky2001fidelity}
$\mathcal{F}(\mathcal{D},\mathcal{U}_{\ir})$
between $\mathcal{D}$ and $\mathcal{U}_{\ir}$
which reads as follows
\begin{align}
  \label{eq:1}
  \mathcal{F}(\mathcal{D},\mathcal{U}_{\ir})
  \coloneqq \frac{1}{d^2_{\ir}}f (D, \KetBra{U_{\ir}}{U_{\ir}} )
\end{align}
where
$f(\rho,\sigma) \coloneqq
\bigl(\Tr\sqrt{\sqrt{\sigma}\rho\sqrt{\sigma}
    }\bigr)^2$ is the state fidelity,
and $D$ and $\KetBra{U_{\ir}}{U_{\ir}}$ are the
Choi
%-Jamilkovski
operators \cite{choi1975completely}
%,jamiolkowski1972linear}
of the channels $\mathcal{D}$ and
$\mathcal{U}_{\ir}$, given by
$D= (\mathcal{D}\otimes \mathcal{I} ) (\KetBra{\mathds{1}}{\mathds{1}})
$ and
$\Ket{U_{\ir}}= (U_{\ir}\otimes \mathds{1}
)\Ket{\mathds{1}}$.
Here, we are using he notation
$\Ket{A}\coloneqq \sum_{i,j=1}^{d_\ir}
A_{i,j}\ket{i}\ket{j}$ for an operator
$A= \sum_{i,j=1}^{d_\ir} A_{i,j}\ketbra{i}{j}$.
If, as in our case, one of the two channels is
unitary, the channel fidelity can be expressed as
the average of the state fidelity over pure input
states, i.e.
$d_{\ir}
\mathcal{F}(\mathcal{D},\mathcal{U}_{\ir})=(d_{\ir}+1)\int d\psi \bra{\psi}
U^{\dag}_{\ir}\mathcal{D}(\ketbra{\psi}{\psi})
U\ket{\psi} - 1$, where $d\psi$ is the Haar
measure of the group $\mathsf{SU}(d_\ir)$.
This criterion also coincides with the one used in ref. 
\cite{Hayden_2022,colla2022open}
to assess the distance between dissipative superoperators.
By inserting Equation \eqref{eq:2} into Equation
\eqref{eq:1} we obtain
\begin{align}
  \label{eq:3}
  &\mathcal{F}(\mathcal{D},\mathcal{U}_{\ir}) =
  \frac{1}{d^2_{\ir}}  \Tr_\uv[O_\uv \rho_\uv O_\uv^\dagger],
  \\
&O_\uv \coloneqq \Tr_\ir[(U_\ir^\dagger\otimes \mathds{1}_\uv)U].\label{eq:4}
\end{align}
We notice that the quantity we want to maximise is invariant under $O_\uv \rightarrow V O_\uv$ for any unitary $V$.

Clearly, when the dynamics factorises as
$U=V_{\ir}\otimes V_{\uv}$, the maximum value
$\mathcal{F}(\mathcal{D},\mathcal{U}_{\ir}) =1$ is
simply achieved by taking $U_\ir=V_\ir$.  On the
other hand, it is worth noticing that the
condition
$\mathcal{F}(\mathcal{D},\mathcal{U}_{\ir}) =1$
does not necessarily imply that $U$ factorizes; an
easy counterexample is a controlled unitary
$U = U^{0}_{\ir} \otimes \ketbra{0}{0} +
U^{1}_{\ir} \otimes \ketbra{1}{1} $ with
$\rho_{\uv} \coloneqq \ketbra{0}{0}$.  However, if
$\rho_{\uv}$ has full rank, then
$\mathcal{F}(\mathcal{D},\mathcal{U}_{\ir}) =1$
does imply that the evolution $U$ factorizes
\footnote{ In general, if $X$ is an operator with
  $0\le X\le \mathds{1}$ and $\rho$ is a density
  matrix, then $\Tr[\rho X]=1$ iff
  $\bra{\lambda} X \ket{\lambda}=1$ for every
  eigenvector $\ket{\lambda}$ of $\rho$ .  In our
  case, one can easily show that
  $X\coloneqq (O_\uv^\dagger O_\uv)/d_\ir^2$
  satisfies $0\le X\le \mathds{1}_\uv$.  If
  $\rho_\uv$ has full rank and $\Tr[\rho X]=1$
  then $X=\mathds{1}_\uv$, which implies that
  $O_\uv/d_\ir$ is unitary.  }.  
  
  \emph{IR-UV
  weakly coupled.} The effective unitary dynamics
will generally be a poor approximation for most
choices of $U$. For instance, a Haar-random
unitary typically generates maximal entanglement
between UV and IR degrees of freedom even after a
single step. However, many physical theories
possess additional structures, such as locality
and symmetries, that set them apart from a typical
Haar-random unitary. From the discussion above, we
understand that unitaries allowing for a
well-approximated effective unitary dynamics are
those that approximately factorize.

Therefore, we now specialise to this  case and
consider
\begin{equation}\label{eq:weak_mix}
U=(V_{\ir}\otimes V_{\uv}) U_{\mix}(\theta),
\end{equation}
where $U_\mix(\theta)$ is analytic at $\theta=0$
with $U_{\mix}(0)=\mathds{1}$.  Then UV and IR are
weakly coupled when $\theta\ll 1$.  The condition
of weak IR-UV coupling is far from artificial; for
example, it is satisfied in quantum field theory,
when one wants to find the effective dynamics of
some light field weakly coupled to a heavy field
\cite{polchinski1992effective, 10.1007/BFb0104294,
  kaplan2005five, agon2018coarse}.  Further below
we will show that it is also applicable to an
interesting example of a quantum walk.

We now proceed to find $U_\ir$ to leading order in
$\theta$.  Due to the invariance of $\mathcal{F}$
under $O_\uv\rightarrow VO_\uv$, we can safely ignore
$V_\uv$.  Then we have
$ O_{\uv} = \Tr_{\ir}[(U_{\ir}^\dagger
V_{\ir}\otimes \mathds{1}_{\uv})
U_{\mix}(\theta)],$ and from the discussion above,
it is clear that $U_{\ir}^\dagger V_{\ir}$ goes to
the identity for $\theta\rightarrow 0$.  This
suggests to expand both $U_\mix$ and
$V_\ir^\dagger U_\ir$ in a Taylor expansion in
$\theta$ and determine $\mathcal{F}$ order by
order.  Namely, let
\begin{equation}\label{eq:u_expansions}
\begin{split}
&U_\mix=\mathds{1}+i\theta H_{\mix} +\theta^2 L_{\mix} + O(\theta^3),\\
&V_\ir^\dagger U_\ir=\mathds{1}+i\theta H_{\ir} +\theta^2 L_{\ir} + O(\theta^3).
\end{split}
\end{equation}
Unitarity implies that $H_{\mix}=H_{\mix}^\dagger$ and $L_{\mix}+L_{\mix}^\dagger=-H_{\mix}^2$, and similarly for $H_{\ir}$ and $L_{\ir}$. 
Plugging these two expansions in \eqref{eq:3}, and after some algebra, we find 
\begin{align}\label{eq:weak_mix_corr}
\begin{split}
&\mathcal{F}=1-\theta^2\bigl(\langle \delta^2\rangle_\ir - \langle \delta\rangle_\ir^2 + \mu(H_{\mix})\bigr),\\
&\delta \coloneqq H_{\ir}-\Tr_\uv[(\rho_\uv\otimes \mathds{1}_\ir) H_{\mix}],
\end{split}
\end{align}
where $\langle \cdot \rangle_\ir = \Tr_\ir[\,\cdot\,]/d_\ir$.
See \cite{supp} for the details of the derivation and for the explicit form of $\mu$.
For the moment, the important point is that $\mu$ does
not depend on $H_{\ir}$.  
The first two terms
in the parentheses on the r.h.s.\ of Eq.\
\eqref{eq:weak_mix_corr} instead do  depend on
$H_{\ir}$ and are clearly non-negative, and so $\mathcal{F}$ is maximised when $\delta=0$, which fixes $H_\ir$.
Exponentiating $H_\ir$ to make sure that $U_\ir$ is unitary, we find  
\begin{equation}\label{eq:pert_rule}
\begin{split}
U_\ir = V_\ir e^{i\theta H_{\ir}},\quad H_{\ir}=\Tr_\uv[(\rho_\uv\otimes \mathds{1}_\ir) H_{\mix}].
\end{split}
\end{equation}
We conclude that the best unitary approximation for the IR dynamics is obtained by taking $H_\ir$ as a mean-field average over $\rho_\uv$. 

To leading order, we also know the error
introduced by the unitarity constraint, i.e.\
$\mu$.  If we decompose the mixing operator as
\begin{align}\label{eq:hmix_dec}
  H_{\mix}=\sum_{\lambda =0}^{{d ^2_\ir} -1}H_{\uv,\lambda} \otimes H_{\ir,\lambda}
\end{align}
where $\{H_{\ir,\lambda}\}_{\lambda =0}^{{d ^2_\ir} -1}$ is a basis of the space of Hermitian operator on  
 $\mathcal{H}_\ir$, and define the connected correlator as
$\langle AB\rangle^c = \langle AB\rangle-\langle A\rangle \langle B\rangle$ 
where $\langle A\rangle = \Tr[A \rho]$, we have \cite{supp}
\begin{align}\label{eq:mu2}
\mu =\sum_{\lambda\lambda'}\langle H_{\ir,\lambda}H_{\ir,\lambda'}\rangle_\ir^c \langle H_{\uv,\lambda}H_{\uv,\lambda'}\rangle_\uv^c.
\end{align}
We can further simplify this expression if we choose a basis such that
$H_{\ir,0} = \mathds{1}_\ir$, and $\Tr[H_{\ir,\lambda}H_{\ir,\lambda'}]
  = d_\ir \delta_{\lambda,\lambda'}$.
In this basis, we find $\langle H_{\ir,\lambda}H_{\ir,\lambda'}\rangle_\ir^c  = 1$ if 
$\lambda = \lambda' \neq 0$ and  0 otherwise, and $\mu$ becomes 
\begin{align}\label{eq:mu3}
\mu =\sum_{\lambda \neq 0 }\langle H_{\uv,\lambda}^2\rangle_\uv^c.
\end{align}
This is simply the sum of energy variances in $\rho_\uv$.
Putting everything together, we find that the best achievable $\mathcal{F}$ is 
\begin{equation}
\mathcal{F}=1-\theta^2\sum_{\lambda\neq0}\langle H_{\uv,\lambda}^2\rangle^{c}_{\uv},
\end{equation}
where the UV Hamiltonians are defined by decomposing $H_\mix$ as explained above.

\emph{Effective dynamics of the Dirac QW.}  In
this section, we apply our prescription to a one
dimensional discrete-time quantum walk given by the unitary operator
\begin{equation}\label{eq:dirac_W}
U =\begin{pmatrix}
\cos(\theta)\,T^\dagger & -i\sin(\theta)\\
-i\sin(\theta) & \cos(\theta)\,T
\end{pmatrix}, \quad
U \in \mathcal{L}(l_2(\mathbb{Z})\otimes \mathbb{C}^2)
\end{equation}
where $T$ is the translation operator, which acts
on the position basis as $T\ket{x}=\ket{x-1}$, and
$\theta\in [0,\pi/4]$ is the only free parameter of the
theory.  The update rule is local, i.e.\
$\psi(x,t)$ only depends on $\psi(y,t-1)$ for
$y\in \{x-1,x,x+1\}$. So this model has a strict
light cone, i.e.\ at each time step information
propagates at most by one lattice point.  This
model is known to reproduce the Dirac equation in
one dimension in the limit of small momenta 
%and
%small $\theta$ 
\cite{Bisio2015244}, and for this
reason it is called Dirac QW.  Given the
translational symmetry of the system, it is
convenient to work in momentum space and the
walk operator becomes
$U = \int_{2\pi}U(p)\otimes \ketbra{p} dp$, where 
\begin{equation}\label{eq:walk_p}
U(p) =\begin{pmatrix}
e^{ip}\cos(\theta) & -i\sin(\theta)\\
-i\sin(\theta) & e^{-ip}\cos(\theta)
\end{pmatrix}.
\end{equation}

\begin{figure}
\centering
\includegraphics[width=0.35\textwidth]{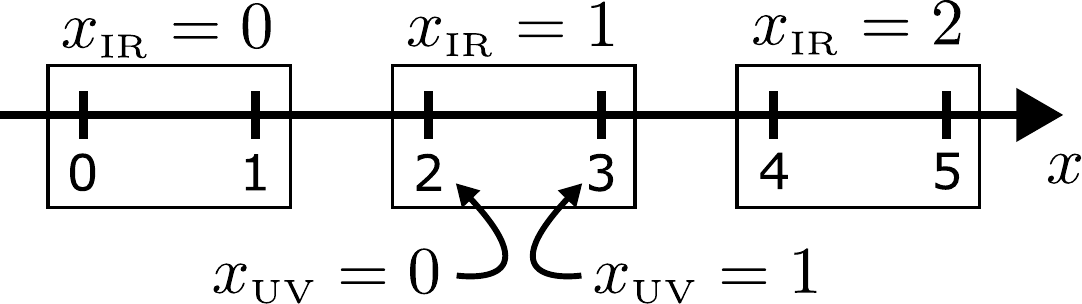}
\caption{Coarse-graining procedure for a QW on $\mathbb{Z}$.}
\label{fig:coarse}
\end{figure}

The lattice suggests a simple way of coarse-graining the Hilbert space. 
We group the sites of the lattice in pairs and relabel
positions by two indices: the first specifying the
bin and the second the position inside the bin, namely
\begin{equation}
\ket{x}=\ket{x_{\ir}}\otimes \ket{x_{\uv}},
\end{equation}
where $x_{\ir}=\lfloor x/2\rfloor\in \mathbb{Z}$
and $x_{\uv}=(x\mod2)\in \{0,1\}$, such that
$x=2x_{\ir}+x_{\uv}$ (see Fig. \ref{fig:coarse}).  
The total Hilbert space
$\mathcal{H}$ is then factorized as
$\mathcal{H} = \mathcal{H}_{\ir} \otimes\mathcal{H}_{\uv}$ where 
$\mathcal{H}_{\ir}=l_2(\mathbb{Z})\otimes
\mathbb{C}^2$ tracks  the bin and the
internal dof of the walker, while
$\mathcal{H}_{\uv}=\mathbb{C}^2$ keeps track of 
whether the walker is on an even or odd site within the bin.  
Correspondingly, in momentum space
$\ket{p}=(2\pi)^{-1/2}\sum_x e^{-ipx}\ket{x}$, we
have that
\begin{equation}
\ket{p}=\ket{2p}_\ir\otimes (\ket{0}_\uv+e^{-ip}\ket{1}_\uv),
\end{equation}
where in the parentheses $\ket{0}_\uv$, $\ket{1}_\uv$ are the UV position basis eigenstates.

We will accompany the spatial coarse-graining with
a temporal one.  The motivation is twofold.  On
one hand, if we assume some experimental
limitations on the spatial resolution, it seems
reasonable to assume that similar limitations will
hold for the temporal resolution of measurements.
On the other hand, this guarantees that the
coarse-graining preserves the light cone. 
A rather straightforward calculation gives
  $U^2 = (V_{\ir}\otimes V_{\uv})U_{\mix}(\theta)$, where
\begin{align}
  \label{eq:5}
  V_{\ir} = \begin{pmatrix}
T_\ir^\dagger & 0\\
0 & T_\ir
\end{pmatrix}, \quad V_{\uv} = I,
\end{align}
$T_\ir$ is the translation operator on
$\mathcal{H}_\ir$ and $U_\mix(\theta)$ is a
unitary operator such that
$U_\mix(0)=\mathds{1}$.
Then, in the limit of interest, i.e.\
$\theta\ll 1$, $U^2$ approximately factorises, and
we can apply Equation \eqref{eq:pert_rule}.  Notice that in
this case, $\mathcal{H}_\ir$ has infinite
dimension, so we would technically be beyond the
validity of the result of the previous section.
We could circumvent this problem by placing the
walk on a finite but sufficiently long chain with
periodic boundary conditions.  This is equivalent
to restrict the set of allowed observables to
those with a sufficiently small support.  In any
case, the formally infinite factors of $d_\ir$
cancel in the formula for the mean field effective
unitary, so to avoid cluttering the notation we
keep the walk on $\mathbb{Z}$.  In \cite{supp}, we
repeat the calculations of this section on a
finite ring and check that the result is indeed
the same.

Applying \eqref{eq:pert_rule} with a generic UV state
$\rho_\uv = (\mathds{1}+r_x\sigma_x + r_y\sigma_y+r_z\sigma_z)/2$ we obtain   $U_\ir = \int_{2\pi}U_\ir(p)\otimes \ketbra{p}_\ir dp$
\begin{align}
  \label{eq:6}
  \begin{aligned}
 & U_\ir(p) =\begin{pmatrix}
e^{ip}\cos(\gamma(p)\theta) & -i\sin(\gamma(p)\theta)\\
-i\sin(\gamma(p)\theta) & e^{-ip}\cos(\gamma(p)\theta)
\end{pmatrix}\\
&\gamma(p)\coloneqq r_{x}(1+\cos(p))-r_y\sin(p)    
  \end{aligned}
\end{align}
Notice that all the dependence on UV data is
contained in $\gamma(p)$.  Comparing this equation
with \eqref{eq:walk_p}, we notice that the
functional shape is the same, but the mass term is
now dependent on $p$,
$\theta\rightarrow \gamma(p)\theta$.  The
dispersion relation which originally was
$\cos(\omega(p))=\cos(p)\cos(\theta)$ is now
$\cos(\omega_\ir(p))=\cos(p)\cos(\gamma(p)\theta)$.
Things further simplify if we consider the small momentum regime, i.e. $p \to 0$.
To leading order we have
$\cos(\omega(p))=1-p^2/2-\theta^2/2$ and
$\cos(\omega_\ir(p))=1-p^2/2-2\theta^2 r_x^2$, and in this limit the effective dynamics
is obtained by a simple rescaling,
$\theta\rightarrow 2\theta r_x$.

To quantify the error introduced by this unitary
approximation, we can use \eqref{eq:mu2}. 
We find that \cite{supp}
\begin{equation}\label{eq:mu_dirac}
\mu = \frac{1}{2}(4-3r_x^2-r_y^2).
\end{equation}
Notice that the error is minimised for $r_x=1$, which corresponds to $\rho_\uv=\ketbra{+}$ with $\ket{+}=(\ket{0}+\ket{1})/\sqrt{2}$, and that no choice of $\rho_\uv$ can make this error vanish. 

\begin{figure}
\begin{center}
\includegraphics[width=0.45\textwidth]{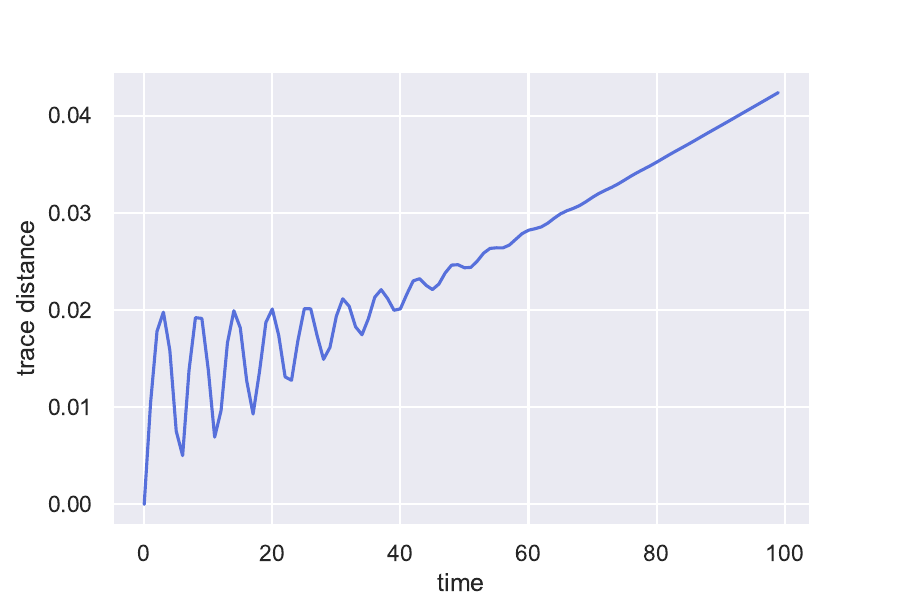}
\caption{Trace distance between coarse-grained dynamics and exact dynamics as a function of time for the propagation of a Gaussian packet.}
\label{fig:gaussian_distance}
\end{center}
\end{figure}

We numerically check the validity of the unitary
approximation by numerically simulating the
propagation of a Gaussian packet.  The result is
displayed in Fig.\ \ref{fig:gaussian_distance}.
After a transient regime with some oscillations, we see that the error grows linearly, due to the accumulation of the small error made at  each step of the evolution.

\emph{Conclusion.}  In this work, we have
considered the question of whether the coarse
grained dynamics of a quantum systems evolving
discretely in time can be effectively described by
a unitary theory.  We have introduced a prescription
for finding the best unitary approximation of the
IR dynamics by maximizing the channel
fidelity. Physically, this can be interpreted as
minimizing dissipation to the unobserved UV dof.
When the UV and IR
degrees of freedom are weakly coupled,
we have carried out the
maximization to leading order in the coupling
constant and we have found that the best unitary
effective dynamics is given by a mean-field-like
average over the UV dof, Eq.\
\eqref{eq:pert_rule}. Our result
agrees with the one found, in the same regime
but for continuous time evolution, in
\cite{agon2018coarse}.
We also showed that the error in the approximation
can be expressed as a sum of energy variances, \eqref{eq:mu3}.
We have then applied these
results to the Dirac QW in the limit of small
masses, $\theta \ll 1$ and small momenta.  We have
found that the effect of the coarse-graining is
to renormalise the mass and we have
checked that this is a good approximation by
numerical simulation of the QW with Gaussain wavepackets.

There are several ways in which the present work
could be further extended.  A natural extension
would be to consider other QW and check whether,
at least in some limit, the coarse-graining
reduces to the renormalisation of some of the
parameters of the walker, as it happened for the
Dirac QW.  In particular, it would be interesting
to consider interacting QW
(e.g. \cite{ahlbrecht2012molecular, Bisio:2017duk}), and see
how the coarse-graining procedure affects the
coupling constant.
In this work we have assumed
that the Hilbert space can be simply factorised
between UV and IR degrees of freedom as
$\mathcal{H}=\mathcal{H}_{\ir}\otimes
\mathcal{H}_{\uv}$, and that the coarse-graining
correspond to tracing-out the UV degrees of
freedom.  It would be interesting to relax both
these assumptions: the first, because it is not
always possible to exactly factorise the Hilbert
space between UV and IR degrees of freedom, the
second, because IR observables don't necessarily
act as the identity on the UV.  In other words, it
would be interesting to consider more
sophisticated ways of splitting the Hilbert space
between IR and UV degrees of freedom, and more
sophisticated coarse-graining procedures. We think
that a good starting point along this road are
\cite{kabernik2018quantum, kabernik2020quantum}
and \cite{zanardi2024operational}.

\acknowledgements
AFR, PP and AB acknowledge financial support from European Union---Next Generation EU through the MUR project Progetti di Ricerca d'Interesse Nazionale (PRIN) DISTRUCT No. P2022T2JZ9.

\bibliographystyle{apsrev4-1}

\bibliography{bibliography}

\clearpage

\section*{\large{SUPPLEMENTAL MATERIAL}}

\section{Derivation of Eq.\ \eqref{eq:weak_mix_corr}}
We derive Eq.\ \eqref{eq:weak_mix_corr} of the main text.
Plugging the expansions \eqref{eq:u_expansions} in \eqref{eq:3}, we find 
\begin{equation}
\mathcal{F}=1-\theta^2\bigl\langle\Tr_\ir\frac{\Delta^2}{d_\ir}-\bigl(\Tr_\ir\frac{\Delta}{d_\ir}\bigr)^2\bigr\rangle_\uv+ O(\theta^3),
\end{equation}
where $\langle \cdot \rangle_\uv = \Tr_\uv[\rho_\uv\,\cdot\,]$ and $\Delta = H_{\mix} - H_{\ir}\otimes \mathds{1}_\uv$. 
One can show that the term in the angle bracket is
negative, as it should since we know that the
whole quantity is upper bounded by 1. 
Some algebra shows that the expression above for $\mathcal{F}$ is equivalent to \ref{eq:weak_mix_corr} for
\begin{align}
  \begin{aligned}\label{eq:mu}
  \mu(H_{\mix})&
        \coloneqq \Tr_\uv[\rho_\uv \Tr_{\ir}\frac{H_{\mix}^2}{d_\ir}]
        \\&-\Tr_\uv[\rho_\uv(\Tr_\ir \frac{H_{\mix}}{d_\ir})^2] + \\
&-\frac{1}{d_\ir}\Tr_{\ir}\bigl(\Tr_\uv[\rho_\uv \otimes \mathds{1}_\ir \cdot H_{\mix}]\bigr)^2 +\\
    &+\frac{1}{d_\ir^2}\bigl(\Tr_\uv[\rho_\uv\Tr_{\ir} H_{\mix}]\bigr)^2.    
  \end{aligned}
\end{align}
Plugging the decomposition \eqref{eq:hmix_dec} in this expression leads to \eqref{eq:mu2}.

\section{Derivation of Eq.\ \eqref{eq:6}}
We derive Eq.\ \eqref{eq:6} from the main text.
As a first step we split $U^2$ in the product of one term which can trivially coarse-grained, and one term that goes to the identity in the $\theta\rightarrow 0$ limit. 
Namely, $U^2 = (\mathds{1}_\uv \otimes V_\ir)\cdot U_{\mix}$, where
\begin{equation}
V_\ir = \begin{pmatrix}
T_\ir^\dagger & 0\\
0 & T_\ir
\end{pmatrix},\quad 
U_\mix = \begin{pmatrix}
A & -iB\\
-iB^\dagger & A^\dagger
\end{pmatrix}
\end{equation}
with 
$A\coloneqq\cos^2(\theta)\mathds{1} -\sin^2(\theta) T^2$ and $B\coloneqq \sin(\theta)\cos(\theta)(T+T^\dagger)T^2$.
Since $U_\mix$ is analytic as a function of $\theta$ and $U_\mix\rightarrow \mathds{1}$ for $\theta\rightarrow 0$, we can apply \eqref{eq:pert_rule}.
Taking the derivative of $U_\mix$ and setting $\theta=0$, we find that $H_\mix = \int_{2\pi} H_\mix(p)\otimes \ketbra{p} dp$ with
\begin{equation}
H_\mix(p)=-2\cos(p)\begin{pmatrix}
0 & e^{-i2p}\\
e^{2ip} & 0
\end{pmatrix}.
\end{equation}
To compute $H_{\ir}=\Tr_\uv[\rho_\uv\otimes \mathds{1}_\ir\cdot H_\mix]$, we use that 
\begin{equation}
\ketbra{p} = \begin{pmatrix}
1 & e^{ip}\\
e^{-ip} & 1
\end{pmatrix}_{\!\uv}\otimes \ketbra{2p}_\ir,
\end{equation}
and paramatrise $\rho_\uv = (\mathds{1}+r_x\sigma_x + r_y\sigma_y+r_z\sigma_z)/2$.
We find that
\begin{equation}
\Tr_\uv\bigl[\rho_\uv \cdot \ketbra{p}\bigr]=(1+r_{xy}(p))\ketbra{2p}_\ir,
\end{equation}
where $r_{xy}(p)\coloneqq r_x\cos(p)-r_y\sin(p)$.
Using this, we find 
\begin{equation*}
H_{\ir}=-2\int_{-\pi}^{\pi}(1+r_{xy}(p))\cos(p)\begin{pmatrix}
0 & e^{-i2p}\\
e^{i2p} & 0
\end{pmatrix}\otimes\ketbra{2p}_\ir.
\end{equation*}
Rescaling the integral $p\rightarrow p/2$ and using $\ket{p\pm 2\pi}=\ket{p}$, we can rewrite the integral as follows
\begin{align}
&H_{\ir}=\gamma(p)\begin{pmatrix}
0 & e^{-ip}\\
e^{ip} & 0
\end{pmatrix}\otimes\ketbra{p}_\ir \\
&\gamma(p) := r_{x}(1+\cos(p))-r_y\sin(p) 
\end{align}
Exponentiating $H_{\ir}$ and multiplying by $V_\ir$, we obtain Equation \eqref{eq:6}. 

\section{Derivation of Eq.\ \eqref{eq:mu_dirac}}
We derive Eq.\ \eqref{eq:mu_dirac} of the main text.
We can decompose $H_{\mix}$ as $H_{\mix}=\int_{2\pi}H_{\uv}(p) \otimes H_{\ir}(p)\, dp$, with
\begin{equation}
H_{\uv}(p) = -2\cos(p/2)\begin{pmatrix}
0 & e^{ip/2}\\
e^{-ip/2} & 0
\end{pmatrix},
\end{equation}
and
\begin{equation}
H_{\ir}(p) = \begin{pmatrix}
0 & e^{-ip}\\
e^{ip} & 0 
\end{pmatrix}\otimes \ketbra{p}_\ir.
\end{equation}
It's not exactly clear how to compute the trace over the IR dof as these are formally infinite. 
In particular, we need to give a prescription for dealing with expressions such as 
$\Tr_\ir[\ketbra{p}_\ir]/{d_\ir}\propto (\sum_{x_\ir}1)/ d_\ir$, 
where both $d_\ir$ and the sum are formally infinite. 
We can deal with this by putting everything in a box, i.e.\ consider a closed chain of size $4L$. Then after coarse-graining we are left with $2L$ sites. 
Taking into account also the coin dof, we have $d_\ir=4L$, and $\sum_{x_\ir}1=2L$, so the two infinities cancel out and we are left with $\Tr_\ir[\ketbra{p}_\ir]/d_\ir=4\pi$.
Using this, we find that 
\begin{equation}
\begin{split}
&\langle H_\uv(p)\rangle = 2\cos(p/2)(r_x\cos(p/2)-r_y\sin(p/2)),\\
&\langle H_\uv(p)H_\uv(q)\rangle = 4\cos(p/2)\cos(q/2),\\
&\phantom{\langle H_\uv(p)T_\uv(q)\rangle}\times\bigl(\cos\frac{p-q}{2}-r_z\sin\frac{p-q}{2}\bigr)\\
&\langle H_\ir(p)\rangle = 0,\quad \langle H_\ir(p)H_\ir(q)\rangle = \frac{1}{2\pi}\delta(p-q),
\end{split}
\end{equation}
which leads to \eqref{eq:mu_dirac}.

\section{Numerics details}\label{app:numerics}
We provide some further details on the numerics performed to generate Fig.\ \ref{fig:gaussian_distance}. 
To simulate the QW we consider a closed chain of size $2L=2000$, and we pick  $\theta=0.2$.
The Gaussian packet is given by
\begin{equation}
\begin{split}
&\ket{\psi_{\sigma_k, k_0, x_0}} \coloneqq\int_{-\pi}^{\pi}dk\, \psi_{\sigma_k, k_0, x_0}(k)\ket{w_+(k)}\otimes \ket{k},\\
&\psi_{\sigma_k, k_0, x_0}(k)\coloneqq \frac{1}{N_{\sigma_k, k_0}}\exp\Bigl[-\frac{(k-k_0)^2}{2\sigma_k^2}+ikx_0\Bigr].
\end{split}
\end{equation}
Here $\ket{w_\pm(k)}$ are the eigenvectors of \eqref{eq:dirac_W}, which are given by  \cite{Bisio:2017duk}
\begin{equation}
\begin{split}
&\ket{w_\pm(p)}=\frac{1}{M_\pm(p)}\begin{pmatrix}
-i\sin(\theta)\\
h_\pm(p)
\end{pmatrix},\\
&h_\pm(p)\coloneqq -i\bigl(\pm\sin\omega(p)+\cos(\theta)\sin(p)\bigr),
\end{split}
\end{equation}
with $\omega(p)\coloneqq \arccos[\cos(\theta)\cos(p)]$ and $M_\pm(p)^2\coloneqq \cos(\theta)^2+\abs{h_\pm(p)}^2$. 
Before continuing considering the numerical simulation, we explain some properties of this state.

We are interested in the $\sigma_k\ll 1$ regime, for which, as we see below, the wavefunction changes sufficiently slowly in the position basis, that the state is approximately factorised between UV and IR. In this regime, we can evaluate the integral over $k$ by saddle approximation and find that in the position basis the packet looks like 
\begin{align}
&\ket{\psi_{\sigma_k, k_0, x_0}}\approx \ket{w_+(k_0)}\otimes \ket{\phi}\\
&\ket{\phi}\coloneqq \sum_x \exp\Bigl[-\frac{(x-x_0)^2}{2\sigma_x^2}-ik_0(x-x_0)\Bigr] \ket{x}\nonumber
\end{align}
where $\sigma_x=1/\sigma_k$, and we have omitted the normalisation. 
Let's focus on $\ket{\phi}$. Under the mapping $\ket{x}\rightarrow \ket{x_\ir}\otimes \ket{x_\uv}$, this becomes
\begin{equation}
\begin{split}
\ket{\phi}&=\sum_{x_\ir}\biggl[ \exp\Bigl[-\frac{2(x_\ir-x_{\ir,0})^2}{\sigma_{x}^2}\Bigr]e^{-i2k_{0}(x_\ir-x_{\ir,0})}\ket{x_\ir}\\
&\phantom{\sum_{x_\ir}\biggl[\quad}\otimes \sum_{x_\uv}\exp\Bigl[-\frac{2(x_\ir-x_{\ir,0})(x_\uv-x_{\uv,0})}{\sigma_x^2}\\
&\phantom{\sum_{x_\ir}\biggl[\quad\otimes \sum_{x_\uv}\exp\Bigl[}-\frac{(x_\uv-x_{\uv,0})^2}{2\sigma_x^2}\Bigr]\\
&\phantom{\sum_{x_\ir}\biggl[\quad}\times e^{-ik_0(x_\uv-x_{\uv,0})}\ket{x_\uv}\biggr],
\end{split}
\end{equation}
where we have set $x_0=2x_{\ir,0}+x_{\uv,0}$. 
For $\sigma_x\gg 1$ we can drop the first exponential inside the sum over $x_\uv$, and we find that $\ket{\phi}$ approximately factorises. Moreover, the UV state is $\rho_\uv=\ketbra{\phi_\uv}$, with
\begin{equation}
\ket{\phi_\uv}=\frac{1}{\sqrt{2}}(\ket{0}+e^{-ik_0}\ket{1}).
\end{equation}
Notice that here we have dropped an irrelevant phase depending on $x_{\uv,0}$.
From the analysis of the main text, we know that in the limit of small momenta all we need to do is renormalise the mass $\theta \rightarrow 2\theta r_x$, and for the staste above $r_x=\cos k_0$.  

In Fig.\ \ref{fig:gaussian_distance}, we plot the trace distance between the IR density matrix obtained by either evolving and then tracing or first tracing and evolving with $U_\ir$, i.e.\ 
\begin{equation}
\mathcal{E}_n = \frac{1}{2}\bigl\lVert \Tr_\uv [U^{2n}\rho (U^{\dagger})^{2n}]- U_\ir^n \rho_\ir (U_\ir^\dagger)^{n}\bigr\rVert_1.
\end{equation}
The plot is generated for a gaussian wavepacket with $\sigma_k=0.02$, $k_0=0.2$, and $x_0=-200$. 

\section{QW in a box}\label{app:box}
We rederive the effective dynamics for the Dirac quantum walk of the main text, regularising the IR by putting it in a box. 
Namely, we consider now a QW on a closed chain of size $4L$. The Hilbert space is spanned by $\ket{x}$ for $x=0, 1, \dots, 4L -1$, and $\ket{x=4L}\equiv \ket{x=0}$. 
Momenta eigenstates are now
\begin{equation}
\ket{p}=\frac{1}{\sqrt{4L}}\sum_{x=0}^{4L-1}e^{-i\frac{\pi p}{2L}x}\ket{x}\,, p\in \mathbb{Z},
\end{equation}
and $\ket{p}=\ket{p+4L}$. 
We take as first Brillouin $p\in [-2L, 2L-1]$. 
The coarse-graining corresponds to 
\begin{equation}\label{eq:box_coarse_grain}
\ket{p}=\frac{1}{\sqrt{2}}\bigl(\ket{0}_\uv + e^{-i\frac{\pi p}{2L}}\ket{1}_\uv\bigr)\otimes \ket{p}_\ir,
\end{equation}
with
\begin{equation}
\ket{p}_\ir=\frac{1}{\sqrt{2L}}\sum_{x_\ir=0}^{2L-1}e^{-i\frac{\pi p}{L}x_\ir}\ket{x_\ir}_\ir\,, p\in \mathbb{Z},
\end{equation}
where $x_\ir = 0, 1,\dots, 2L-1$ and $\ket{p}_\ir=\ket{p+2L}_\ir$. 
We take as first Brillouin zone $p\in [-L, L-1]$. 

Following the same steps as for the infinite case, we find that 
\begin{equation}
H_\mix=-2\sum_{p=-2L}^{2L}\cos(\frac{\pi p}{2L})\begin{pmatrix}
0 & e^{-i\frac{\pi p}{L}}\\
e^{i\frac{\pi p}{L}} & 0
\end{pmatrix}\otimes \ketbra{p}.
\end{equation}
Using \eqref{eq:box_coarse_grain} and rearranging the summation indices we arrive to 
\begin{equation}
\begin{split}
H_\mix&=-2\sum_{p=-L}^{L-1}\cos(\frac{\pi p}{2L})
\begin{pmatrix}
0 & e^{-i\frac{\pi p}{2L}}\\
e^{i\frac{\pi p}{2L}} & 0
\end{pmatrix}_{\!\uv}\\
&\otimes \begin{pmatrix}
0 & e^{-i\frac{\pi p}{L}}\\
e^{i\frac{\pi p}{L}} & 0
\end{pmatrix}_{\!c}\otimes \ketbra{p}_\ir,
\end{split}
\end{equation}
where the subscript $c$ denotes the coin factor. 
Computing $H_{\ir}=\Tr_\uv[\rho_\uv\otimes \mathds{1}_\ir\cdot H_\mix]$ we find
\begin{equation}
\begin{split}
&H_{\ir}=-i\sum_{p=-L}^{L-1}r_{xy}(p)
\begin{pmatrix}
0 & e^{-i\frac{\pi p}{L}}\\
e^{i\frac{\pi p}{L}} & 0
\end{pmatrix}_{\! c}\
\otimes\ketbra{p}_\ir\\
&r_{xy}(p)\coloneqq r_x\bigl(1+\cos\frac{\pi p}{L}\bigr)-r_y \sin\frac{\pi p}{L}
\end{split}
\end{equation}
From here on the calculation follows identically to the infinite case and we find
$\theta_\ir(p) = [r_x\bigl(1+\cos(\pi p/L)\bigr)-r_y \sin(\pi p/L)]\theta$. 

To compute $\mu$ we parametrise $\rho_\uv$ as explained in the main text and we take $\rho_\ir = \mathds{1}_\ir/d_\ir$ which is now well defined with $d_\ir = 4L$ (which includes a factor $2$ for the coin space). 
To compute $\mu$ use Eq.\ \eqref{eq:mu}, we find
\begin{align}
\begin{aligned}
&\Tr_\uv\Bigl[\rho_\uv \Tr_{\ir}\frac{H_{\mix}^2}{d_\ir}\Bigr]=
-\frac{2}{L}\sum_{p=-L}^{L-1}\cos^2\Bigl(\frac{\pi p}{2L}\Bigr)\\
&\Tr_\uv\Bigl[\rho_\uv\Bigl(\Tr_\ir \frac{H_{\mix}}{d_\ir}\Bigr)^2\Bigr]=0 \\
&\frac{1}{d_\ir}\Tr_{\ir}\bigl(\Tr_\uv[\rho_\uv \otimes \mathds{1}_\ir \cdot H_{\mix}]\bigr)^2=
\frac{-1}{2L}\sum_{p=-L}^{L-1}r_{xy}^2(p)\\
&\frac{1}{d_\ir^2}\bigl(\Tr_\uv[\rho_\uv\Tr_{\ir} H_{\mix}]\bigr)^2=0.    
\end{aligned}
\end{align}
The sums above can be calculated analytically,
\begin{align}
\begin{aligned}
&\frac{2}{L}\sum_{p=-L}^{L-1}\cos^2\Bigl(\frac{\pi p}{2L}\Bigr)=2\\
&\frac{1}{2L}\sum_{p=-L}^{L-1}r_{xy}^2(p)=\frac{1}{2}(3r_x^2+r_y^2).   
\end{aligned}
\end{align}
So finally we arrive again to \eqref{eq:mu_dirac}.

\end{document}